\documentstyle[psfig]{mn}

\def\etal{{\rm et~al.\ }}

\def\HII{H\,{\sc ii}}
\def\kms{km~s$^{-1}$}

\begin{document}
\label{firstpage}
\title[Limits on radio emission from pulsar wind nebulae]
{Limits on radio emission from pulsar wind nebulae}
\author[B. Gaensler \etal]
{B. M. Gaensler,$^1$\thanks{Hubble Fellow} 
B. W. Stappers,$^2$ 
D. A. Frail,$^3$
D. A. Moffett,$^4$\thanks{Current address: Physics
Department, Furman University, 3300
Pointsett Hwy, Greenville, SC 29613, USA} \newauthor
S. Johnston$^5$ and S. Chatterjee$^{6,3}$ \\
$^1$Center for Space Research, Massachusetts Institute of Technology,
70 Vassar Street, Cambridge, MA 02139, USA; bmg@space.mit.edu \\
$^2$Astronomical Institute ``Anton Pannekoek'', Kruislaan
403, 1098 SJ Amsterdam, The Netherlands \\
$^3$National Radio Astronomy Observatory, P.O. Box 0, Socorro, NM 87801, USA \\
$^4$Physics Department, University of Tasmania, GPO Box
252-21, Hobart, Tasmania 7001, Australia \\
$^5$Research Centre for Theoretical Astrophysics,
University of Sydney, NSW 2006, Australia \\
$^6$Department of Astronomy and Space Sciences, 
Cornell University, Ithaca, NY 14853, USA }

\pagerange{\pageref{firstpage}--\pageref{lastpage}}
\pubyear{2000}

\maketitle
\begin{abstract}

We report on a sensitive survey for radio pulsar wind nebulae (PWN)
towards 27 energetic and/or high velocity pulsars. Observations were
carried out at 1.4~GHz using the Very Large Array and the Australia
Telescope Compact Array, and utilised pulsar-gating to search for
off-pulse emission.  These observing parameters resulted in a considerably
more sensitive search
than previous surveys, and could
detect PWN over a much wider range of spatial scales (and hence ambient
densities and pulsar velocities).  However, no emission clearly
corresponding to a PWN was discovered. Based on these non-detections we
argue that the young and energetic pulsars in our sample have winds
typical of young pulsars, but produce unobservable PWN because they
reside in low density ($n \sim 0.003$~cm$^{-3}$) regions of the ISM.
However, non-detections of PWN around older and less energetic pulsars
can only be explained if the radio luminosity of their winds is less
than $10^{-5}$ of their spin-down luminosity, implying an efficiency
at least an order of magnitude smaller than that seen for young pulsars.

\end{abstract}


\section{Introduction}
\label{sec_intro}

Almost all radio pulsars have rotational periods which are steadily increasing
with time. This spin down corresponds to a loss of rotational kinetic energy
$\dot{E} \equiv 4\pi^2I\dot{P}/ P^3$, where $I$ is the moment of inertia of
the neutron star (assumed to be $10^{45}$~g~cm$^2$) and $P$ is its period;
for the known pulsar population $\dot{E}$ falls in the range
$10^{28}-10^{39}$~erg~s$^{-1}$. The bolometric luminosity of the radio pulses
themselves is in all cases a tiny fraction of the spin-down luminosity
$\dot{E}$, and it is thought that most of the pulsar's
spin-down luminosity is dissipated via a
magnetised wind populated by relativistic electrons and positrons
\cite{rg74,mic82,kc84}.  However it is still not well understood how this
wind is produced, how it evolves as it flows away from the pulsar, what it
is composed of, how its properties depend on those of the pulsar itself, or
how it changes as the pulsar ages.

Particles in the wind move along magnetic field lines
as they stream away from the pulsar magnetosphere, and produce no observable
emission. At some distance from the pulsar, the pressure of the wind is
eventually balanced by an external pressure, and the resulting shock
randomises the pitch angles of the relativistic particles.  These particles
consequently gyrate in the local magnetic field and produce synchrotron
emission.  The properties of the
resulting {\em pulsar wind nebula}\ (PWN) can then be used to determine
various parameters of the pulsar wind which produces it. At radio
wavelengths, PWN are characterised by an amorphous or filled-centre
morphology, a moderate degree of linear polarization ($\sim$20\%), and
relatively flat spectra ($\alpha \sim -0.3$, $S_\nu \propto
\nu^\alpha$) \cite{wp78}.

Various types of PWN are produced, depending on the source of
confinement of the wind. Young pulsars are often still located inside their
associated supernova remnants (SNRs), and the hot gas produced by the SNR
blast-wave provides the confining pressure.  These PWN, also known as
``plerions'', are typified by the Crab Nebula. If the SNR has dissipated,
the confining pressure is then that of
the ambient interstellar medium (ISM). This results in much larger ``ghost
remnants'', which have been proposed but not observed
\cite{bopr73,ccgm83}. In cases where a pulsar has a high space velocity, the
ram pressure resulting from its motion can dominate the ambient gas
pressure, resulting in a bow-shock PWN (e.g.\ Frail \& Kulkarni
1991\nocite{fk91}). 


While PWN can tell us much about pulsar winds, the number of sources we have
to study is small -- at radio wavelengths, fewer than 10 pulsars have
observable PWN. All of these pulsars are very young and have high values of
$\dot{E}$. While $\sim$20\% of SNRs have a ``plerionic'' component, in most
cases the associated pulsar has not been detected, and without knowing the
pulsar parameters it is difficult to constrain the properties of the PWN and
the corresponding wind. Thus it is of considerable interest to target
candidate pulsars with the intention of either finding new PWN, or
determining upper limits on such emission.

Many searches for radio PWN around energetic or fast-moving pulsars
have been carried out, at varying resolutions and surface-brightness
sensitivities, but usually with no success  (e.g. Sch\"{o}nhardt
1974\nocite{sch74}; Weiler, Goss \& Schwarz 1974\nocite{wgs74}; Cohen
\etal\ 1983\nocite{ccgm83}).  The most recent and comprehensive of these
searches, and the only one to specifically target young, energetic or
high velocity pulsars, was the recent survey of Frail \& Scharringhausen
\shortcite{fs97}, hereafter FS97. FS97 imaged regions around 35 pulsars
with the Very Large Array (VLA) at 8.4~GHz, and found no nebular emission
associated with any of their targets. Their stringent upper limits allowed
them to conclude that most pulsars put less than $10^{-6}$ of $\dot{E}$
into radio emission from a PWN, $\sim100$ times less than observed for
the young, high $\dot{E}$, pulsars which power detected radio nebulae.
Based on this result, FS97 concluded
that pulsar winds change in some way as pulsars age and slow down, 
such that they are no longer efficient at producing radio emission.

However, despite these apparently constraining limits, in hindsight the
observing parameters for this search were probably not ideal for looking
for PWN.  First, FS97 argued that for ambient densities of 1~cm$^{-3}$
and pulsar velocities of 150~\kms, PWN around almost all their sources
were likely to be unresolved, even at their high spatial resolution
of $0\farcs8$. However, other choices of ambient density and pulsar
velocity can produce PWN with much larger angular extents, resulting in
a flux density limit much poorer than estimated by FS97.  Furthermore,
the maximum scale to which FS97 were sensitive was only $20''$; they
could not detect PWN larger than this at any flux density. Secondly,
in 40\% of their sample FS97 detected a point source at the position
of the pulsar, but had no way of distinguishing between the compact
PWN they were looking for and the pulsars themselves. By extrapolating
pulsar flux densities from much lower frequencies, FS97 argued that the
flux densities they were detecting at 8.4~GHz were consistent with the
expected pulsed fluxes, and hence concluded that they were not detecting
any PWN in these data.  Finally, a typical PWN (spectral index $\alpha =
-0.3$) has a significantly lower flux density at their observing frequency
of 8.4~GHz than at lower frequencies.

Motivated by these points, we have undertaken an extensive survey
for PWN at both northern and southern declinations, with observing
parameters chosen to give much greater sensitivity to PWN.  First,
we have observed at 1.4~GHz, at which frequency PWN can be expected to
be $\sim$70\% brighter than at 8.4~GHz. Secondly, we have observed at a
reduced spatial resolution of $\sim$12$''$, to give better sensitivity
to extended structure. Thirdly, our observations are in telescope
configurations whose shortest spacings correspond to a spatial scale of
many arcmin. Finally, and most importantly, all our observations have
employed pulsar-gating, in which images are made from data taken only
when the pulsar is off; we are thus sensitive to compact and unresolved
PWN, which might otherwise be masked by the pulsars themselves.

Initial results from this survey, using the Australia Telescope Compact
Array (ATCA), have been presented in two previous papers. Gaensler
\etal\ \shortcite{gsfj98}, hereafter GSFJ98, reported the discovery
of a faint PWN associated with PSR~B0906--49, while Stappers,
Gaensler \& Johnston \shortcite{sgj99}, hereafter SGJ99, presented
non-detections of PWN towards four pulsars.
We here report on the remainder of this survey, consisting of 1.4~GHz
pulsar-gated observations of 27 more pulsars, using the VLA and
ATCA. In \S\ref{sec_obs}, we describe our observations and analysis,
while in \S\ref{sec_results} we present non-detections of PWN towards
these sources. In \S\ref{sec_discuss} we quantify the improvement in
sensitivity of the current survey, and discuss the constraints we can
put on the radio luminosities of pulsar winds from our data.

\section{Observations and Reduction}
\label{sec_obs}

The 27 pulsars observed are listed in Table~\ref{tab_obs}; all were chosen
for their high $\dot{E}$ and/or space velocity. From this sample 22 pulsars
were observed with the VLA \cite{nte83}, while the remaining 5 pulsars were
observed with the ATCA \cite{fbw92}. All observations were carried out at
frequencies near 1.4~GHz.

\begin{table*}
\begin{minipage}{180mm}
\caption{Pulsars surveyed for PWN. Uncertainties in the last
digit are indicated (or are omitted when the uncertainty
is less than one in the last digit); positional uncertainties do
not include systematic errors due to calibration. Three pulsars
were not detected, for reasons discussed in the text. PSR~B1706--44
was included as a test case.}

\label{tab_obs}
\begin{tabular}{llcllccccl} 
Pulsar &  Date & Telescope & \multicolumn{2}{c}{Pulsar Position (J2000)} & 
Pulsar Flux Density & Off-pulse RMS 
&  $\theta_{\rm min}$ & $\theta_{\rm max}$  \\
      & Observed &         &   RA & Dec           & at 1.4~GHz (mJy) &       (mJy~beam$^{-1})$& ($''$) & ($'$) \\
B0114+58 & 1999 Jan 05 & VLA & $\ldots$ & $\ldots$ & $\ldots$ &  0.47 & $19\times10$ & 15 \\
B0136+57 & 1999 Jan 05 & VLA & 01:39:19.76(1) & +58:14:31.68(7) &4.8(4) & 0.30 & $18\times10$ & 15 \\
B0355+54 & 1999 Jan 05 & VLA & 03:58:53.680(5) & +54:13:13.63(5) & 11(2) & 0.26 & $15\times10$ & 15 \\
J0538+2817 & 1999 Jan 05 & VLA & 05:38:25.090 & +28:17:09.41(1) & 4.1(2)&0.40 & $12\times11$ & 15 \\
B0540+23 & 1999 Jan 05 & VLA & 05:43:09.665 & +23:29:06.167(1) & 32.1(1)&0.42 & $12\times11$ & 15 \\
B0611+22 & 1999 Jan 05 & VLA & 06:14:17.020 & +22:29:56.680(5) & 6.9(4)& 0.40 & $12\times11$ & 15 \\
J0631+1036 & 1999 Jan 05 & VLA & $\ldots$ & $\ldots$ & $\ldots$ & 1.8 & $13\times12$ & 15 \\
B0656+14 & 1999 Jan 05 & VLA & 06:59:48.13(2) & +14:14:21.0(4) & 1.5(2) & 0.26 & $12\times12$ & 15 \\
B0736--40 & 1999 Feb 21 & ATCA & 07:38:32.342 & --40:42:40.16 & 87(5) & 0.19 & $17\times8$ & 4.5 \\
B0740--28 & 1999 Jan 05 & VLA & 07:42:49.038 & --28:22:43.331(2) & 25.7(2)&0.53 & $30\times10$ & 15 \\
B1356--60 & 1999 Mar 10 & ATCA & 13:59:58.5(6) & --60:38:07.73(9)& 12.5(3) & 0.29 & $16\times10$ & 4.5 \\
B1449-64 & 1999 Feb 19 & ATCA & 14:53:32.712(1) & --64:13:15.51(1) & 19.5(5) & 0.07 & $11\times9$ & 4.5 \\
B1508--57 & 1999 Feb 18 & ATCA & 15:12:43.041(3) & --57:59:59.94(3) & 7.3(7)  & 0.21 & $10\times9$ & 1.8  \\
B1634--45 & 1999 Feb 21 & ATCA & 16:37:58.729(9) & --45:53:26.7(2) & 1.4(3) &0.16 &  $15\times9$ & 4.5  \\
B1706--16 & 1999 Feb 02 & VLA & 17:09:26.44(1) & --16:40:57.4(3) & 6.6(3) &0.50 & $20\times12$ & 15 \\
B1706--44 & 1999 Feb 02 & VLA & 17:09:42.52(2) & --44:29:06(1) & 7.3(7) & 0.36 & $91\times10$ & 15 \\
B1718--35 & 1999 Feb 02 & VLA & 17:21:32.71(3) & --35:32:46.4(7) & 5(5) & 1.6& $51\times11$ & 15 \\
B1719--37 & 1999 Feb 02 & VLA & 17:22:59.04(7) & --37:11:57(2) & 1.2(2) &0.53 & $51\times11$ & 15 \\
B1727--33 & 1999 Feb 02 & VLA & $\ldots$ & $\ldots$ & $\ldots$ & 1.6 & $40\times9$ & 15  \\
B1730--37 & 1999 Feb 02 & VLA & 17:33:26.74(3) & --37:16:56(1) & 2(1)& 1.2 &$45\times12$ & 15 \\
B1754--24 & 1999 Feb 02 & VLA & 17:57:29.371(1) & --24:22:02.22(2) & 2.8(1)&0.15 & $25\times10$ & 15 \\
B1821--19 & 1999 Feb 02 & VLA & 18:24:00.460 & --19:45:53.571(2) & 8.1(3)& 0.19& $21\times11$ & 15 \\
B1823--13 & 1999 Feb 02 & VLA & 18:26:13.16(3) & --13:34:49.9(7) & 2.2(2) & 0.32 & $16\times12$ & 15 \\
J1835--1106 & 1999 Feb 02 & VLA & 18:35:18.32(2) &  --11:06:16.6(4) & 3(2) & 0.58 & $16\times12$ & 15 \\
B1930+22 & 1999 Feb 02 & VLA & 19:32:22.63(1) & +22:20:53.3(2) & 1.6(2) & 0.54 &$12\times11$  & 15 \\
B1933+16 & 1999 Feb 02 & VLA & 19:35:47.830 & +16:16:39.806(1) & 75(1) & 0.34& $13\times11$ & 15 \\
B2011+38 & 1999 Feb 02 & VLA & 20:13:10.341(1) & +38:45:43.225(7) & 11(1) & 0.82 & $12\times10$ & 15 \\
B2148+63 & 1999 Feb 02 & VLA & 21:49:58.71(3) & +63:29:44.9(2) & 2.9(2) & 0.59 & $14\times10$ & 15 \\
\end{tabular}
\end{minipage}
\end{table*}

The VLA observations were made in the C configuration, using a bandwidth of
25~MHz for $00^{\rm h} < {\rm RA} < 12^{\rm h}$ and 12.5~MHz otherwise, and
with a phase centre corresponding to the catalogued pulsar position.
The observing time for each pulsar
was typically 15~min.  Amplitudes were calibrated using
observations of 3C~286 and 3C~48, assuming 1.4~GHz flux densities of 14.9~Jy
and 16.3~Jy respectively (where 1~Jy~$=10^{-26}$~W~m$^{-2}$~Hz$^{-1}$).
ATCA observations were made in the 6C configuration, using a bandwidth of
128 MHz (further subdivided into 32 spectral channels); amplitude
calibration was carried out using PKS~B1934--638 and assuming a 1.4~GHz flux
density of 14.9~Jy.  For ATCA observations, the phase centre
was offset from the pulsar's catalogued position
by $\sim1'$; each pulsar was observed for approximately
12~hr. Antenna gains and instrumental polarization were
calibrated using observations of strong unresolved sources in the vicinity
of each pulsar; all four Stokes parameters were recorded.

All observations were gated at the pulsar period in order to look for
off-pulse emission at the pulsar position, using ephemerides 
supplied by A.G. Lyne. For the VLA, gating was carried
out by phasing up the array on a nearby calibrator, then integrating on the
pulsar for a few minutes. The analogue sum of the signals from all antennas
was formed from these data, and then folded at the apparent pulse period to
give an un-dedispersed pulse profile. A gate was then set on-line, such that
one IF recorded on-pulse data while the other recorded off-pulse data,
effectively giving two bins of possibly uneven size. The smearing due to
dispersion across the band was sufficiently small for all
pulsars that it was always possible to completely separate on-
and off-pulse emission when choosing the gate.
For the ATCA, visibilities were recorded at high time-resolution
(typically 32 bins per period), and then folded at the apparent pulse period
before being written to disk. Dedispersion (of 32 channels across the 128~MHz
bandwidth) was carried out during data reduction, and
appropriate phase bins were then chosen to generate on- and off-pulse
images.

Data were edited and calibrated using the {\tt MIRIAD}\ and {\tt AIPS}\
packages according to standard procedures \cite{gre96c,sk98}.  On- and
off-pulse images of a field containing each pulsar were formed using
uniform weighting. Each image was deconvolved using either the {\tt
CLEAN}\ algorithm (for fields containing primarily point sources) or a
maximum entropy algorithm (for fields containing significant extended
emission), and then smoothed with a Gaussian restoring beam.  For some
sources the region was significantly confused by extended structure;
in these cases, we constrained the deconvolution process by generating
{\tt CLEAN} boxes from lower resolution Molonglo Galactic
Plane Survey (MGPS) or Northern VLA Sky Survey (NVSS) data \cite{gcl98,ccg+98}.

For each source, the position of the pulsar was determined by fitting in the
$u-v$ plane to the difference of the on- and off-pulse data.  For VLA data,
a flux density for each pulsar was determined by measuring a flux in the
on-pulse image, then scaling using the width of the gate used. For ATCA
data, flux densities were measured directly from ungated data.  The
sensitivity of each image was determined by measuring the off-pulse RMS 
at the pulsar position in each case.

\section{Results}
\label{sec_results}

The positions, fluxes and off-pulse sensitivities for the 27 pulsars
observed are listed in Table~\ref{tab_obs}, along with the range of spatial
scales to which each image was sensitive. 
In most cases, the pulsar was clearly detected in the on-pulse
image, and was completely gated out in the off-pulse image.

As a test of our sensitivity we included PSR~B1706--44 in our VLA
observations, a pulsar which is known to be embedded in a candidate PWN
\cite{fgw94}.  This nebula was easily detected in our image, with a surface
brightness and spatial extent similar to that obtained in previous data.

All but three of the pulsars in our sample were detected in these
observations. For detections the measured interferometric position was
compared with the timing position given in the
pulsar catalogue \cite{tml93}. Our gated position for PSR B1706--44
differs significantly from the original timing position \cite{jlm+92},
but is in reasonable agreement with the interferometric measurements of
Frail, Goss \& Whiteoak 
\shortcite{fgw94} and the new timing position of Wang \etal\
\shortcite{wmp+99}. Our position for PSR B1754--24 has significantly
smaller uncertainties than the catalogued
position.  The error in the latter
is certainly much greater than the beamwidth in the observations
of Kijak \etal\ \shortcite{kkwj98}, and can account for their 4.9~GHz
non-detection
of this flat-spectrum pulsar. The majority of the remaining
positions are consistent, to within $3\sigma$, with positions 
determined from pulsar timing.

For the majority of sources in our sample, the off-pulse image showed no
emission, extended or point-like, at or near the position of the
pulsar. Sources of note are discussed individually below:

\noindent {\em B0114+58:} On-pulse data were corrupted for this source
by hardware problems during the observations, and so no detection of
the pulsar was made. The off-pulse image was uncorrupted, and shows no
emission at the catalogued pulsar position.

\noindent {\em B0136+57:} An unresolved source of flux density
$2.6\pm0.3$~mJy, approximately half of the observed pulsar flux density,
is seen at the pulsar's position in an off-pulse image. This source
is $\sim90$\% linearly polarized, which is similar to the degree of
polarisation seen for the pulsar \cite{gl98}.  Thus the source probably
corresponds to a component of the pulse-profile which was not properly
gated out when the gate was set during observations.


\noindent {\em J0631+1036:} Images of the region were badly corrupted
by sidelobes from the source 4C+10.20 (flux density
2.5~Jy), $19'$ distant. The pulsar was not detected, its tabulated 1.4~GHz flux
density of 0.8~mJy being below the sensitivity of the data. No
other emission at or near the pulsar's position was detected, down to
the sensitivity limit.

\noindent {\em B0656+14:} No off-pulse emission is apparent at the
pulsar's position, but an extended, polarized source is seen $\sim2'$
south of the pulsar, which Cordova \etal\ \shortcite{cmhm89} argue
is possibly associated with the pulsar. We note that we
see no connecting structure between the pulsar
and this source, despite being more sensitive to extended emission than
Cordova \etal\ \shortcite{cmhm89}. It thus seems likely
that this source is unrelated to the pulsar. We note that
an X-ray PWN associated with this source was claimed by 
Kawai \& Tamura \shortcite{kt96}, but has been discredited
by higher resolution data \cite{bkbm99}.

\begin{figure*}
\begin{minipage}{160mm}
\centerline{\psfig{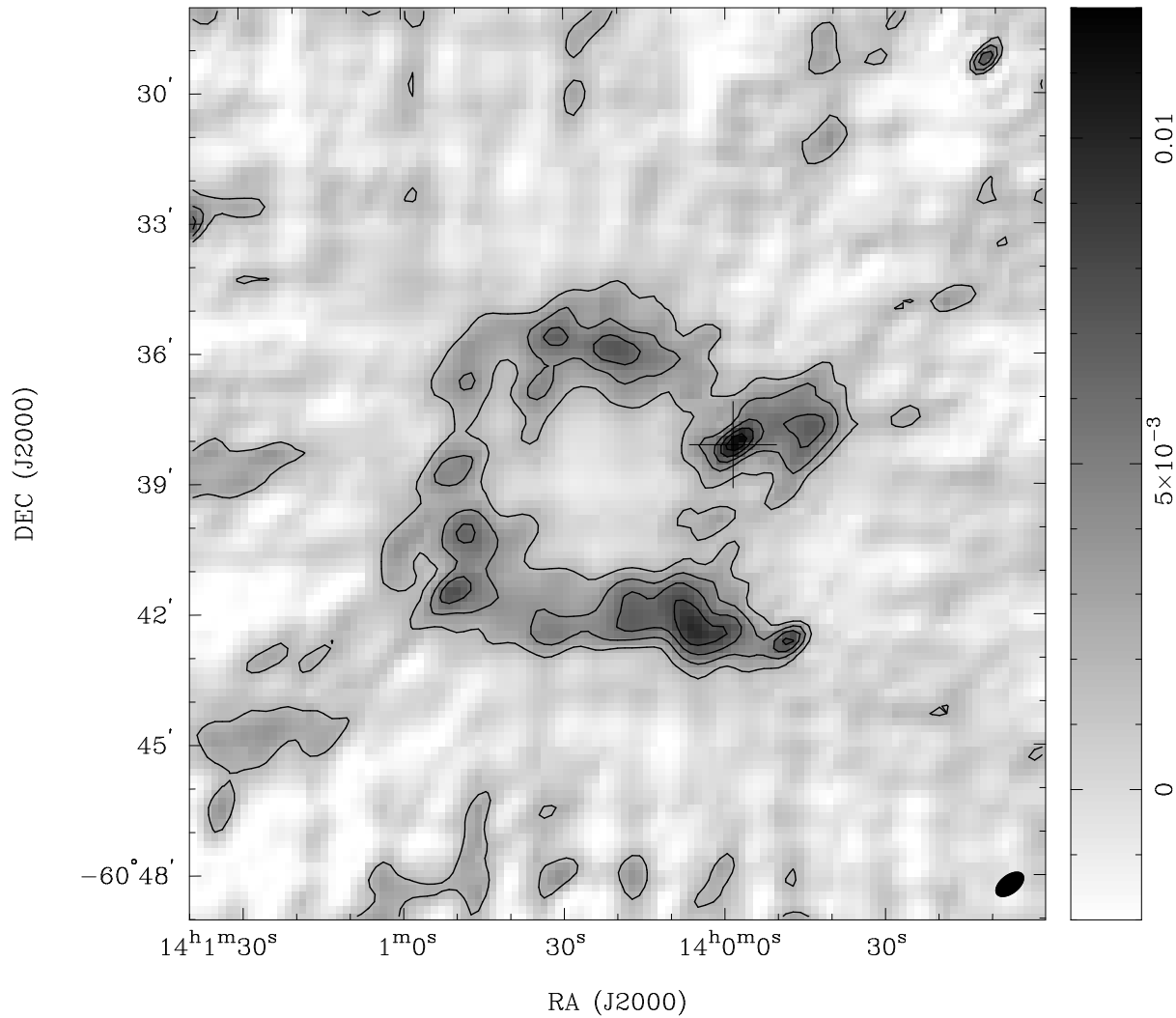}}
\caption{1.4~GHz ATCA image of the region surrounding PSR~B1356--60.
Contours are at levels of 2, 4, 6, 8, 10 and 12~mJy~beam$^{-1}$,
at a resolution of $46''\times26''$ (FWHM shown at lower right).
The pulsar (which has not been gated out in this image) is marked
with a cross, and can be
seen sitting on the western rim of the shell G311.28+1.09.}
\label{fig_b1356}
\end{minipage}
\end{figure*}

\noindent{\em B1356--60:} The region around this pulsar is shown in
Fig~\ref{fig_b1356} -- the pulsar lies on the western rim of a shell of
radio emission. This shell, which we designate G311.28+1.09, is
approximately circular, with a diameter $\sim$9~arcmin and a flux density
$0.04\pm0.01$~Jy at 1.4~GHz. In linear polarization there is
significant confusion from other sources in the region, and so it is not
possible to determine whether the shell is polarized. Based on
morphology alone, we thus consider G311.28+1.09 to be a possible new
SNR, although further observations will be required to confirm this.
However, given the 300-kyr characteristic age of the pulsar, it is
unlikely that there is any physical association between PSR B1356--60
and G311.28+1.09.  In an ungated image, no PWN is apparent around the
pulsar, although the sensitivity to such a source is poor
due to the presence of  G311.28+1.09.

\noindent{\em B1508--57:} This pulsar is in a confused region of the
Galactic Plane. We were forced to discard
short $u-v$ spacings in order
to image the region, limiting the largest spatial scale to which these
observations were sensitive to only $1\farcm8$. No off-pulse emission
was observed, subject to this constraint.

\noindent{\em B1634--45:} The pulse profile for this source reveals an
interpulse separated in phase by 180 degrees from the main pulse, a result
which has been confirmed by recent timing observations (F. Crawford,
private communication).  Gating out both the pulse and interpulse reveals
an unresolved off-pulse source at the pulsar's position of flux density
$0.8\pm0.4$~mJy. This source is $\sim$80\% linearly polarized, similar
to that measured for the main pulsed component.  Since no PWN has ever
been observed to be so highly polarized, we think it unlikely that this
off-pulse source corresponds to an extended nebula; it is more likely
that this emission comes from the pulsar itself.  While we cannot rule out
an error in the gating hardware or software, we note that no other gated
ATCA observations have shown such an effect (see SGJ99).  Alternatives are
that there is a low-level bridge of pulsed emission connecting the two
main components of the pulse profile (cf. PSR~B1259--63; Manchester
\& Johnston 1995\nocite{mj95}), or that the pulse profile contains
an underlying unpulsed component (cf. PSR~J0218+4232; Navarro \etal\
1995\nocite{nbf+95}).  We are planning further ATCA observations of this
source in order to distinguish between these possibilities.

\noindent {\em B1706--16:} An unresolved off-pulse source of flux density
$4.0\pm0.3$~mJy is seen at the pulsar's position. This source is less
than 15\% linearly or circularly polarized, but so is the pulsar itself
\cite{gl98}. While the pulse profile is quite narrow and shows no evidence
for an interpulse, the fact that the VLA gating is set on-line means that,
as for PSR~B0136+57, we are unable to rule out a component of the pulse
profile as the source of this detection. As for PSR~B1634--45 above,
we plan to re-observe this pulsar with the ATCA in order to clarify
this situation.


\noindent {\em B1718--35:} This pulsar is in a complicated region, and
suffers significant confusion from the nearby star-forming region NGC~6334
(e.g.\ Brooks \& Whiteoak 2000\nocite{bw99b}). Gating shows the pulsar
to be located at the center of a $4\arcmin$ radio nebula, G351.70+0.66,
which is also clearly visible in data from the MGPS \cite{gcl98}. This
region has a distinct counterpart at 60~$\mu$m in {\em IRAS}\ data,
and is probably thermal.

\noindent {\em B1727--33:} Two ultra-compact \HII\ regions are in the
field, one of which, G354.19--0.06 \cite{bwhz94}, has a flux density of
0.3~Jy and is located $\sim10'$ from the pulsar position. The pulsar
was not detected, its catalogued flux of 2.9\,mJy corresponding to a
signal-to-noise of only $2\sigma$.

\noindent {\em B1730--37:} The sensitivity of the observations was
reduced by the presence of PMN~J1733--3722 (flux density 0.6 Jy),
just $2'$ away.

\noindent {\em B1754--24:} The $1\sigma$ uncertainty in the right
ascension of this pulsar was previously $14'$; as discussed above, we
have now greatly improved on this position. This more precise position
puts the pulsar  along the same line of sight as the large diffuse \HII\
region G5.33+0.08 \cite{lph96}. Emission from the latter is clearly
seen in NVSS data and in the 90~cm image of Frail, Kassim \& Weiler
\shortcite{fkw94}, but is largely resolved out by our observations.


\section{Discussion}
\label{sec_discuss}

\begin{figure*}
\begin{minipage}{160mm}
\centerline{\psfig{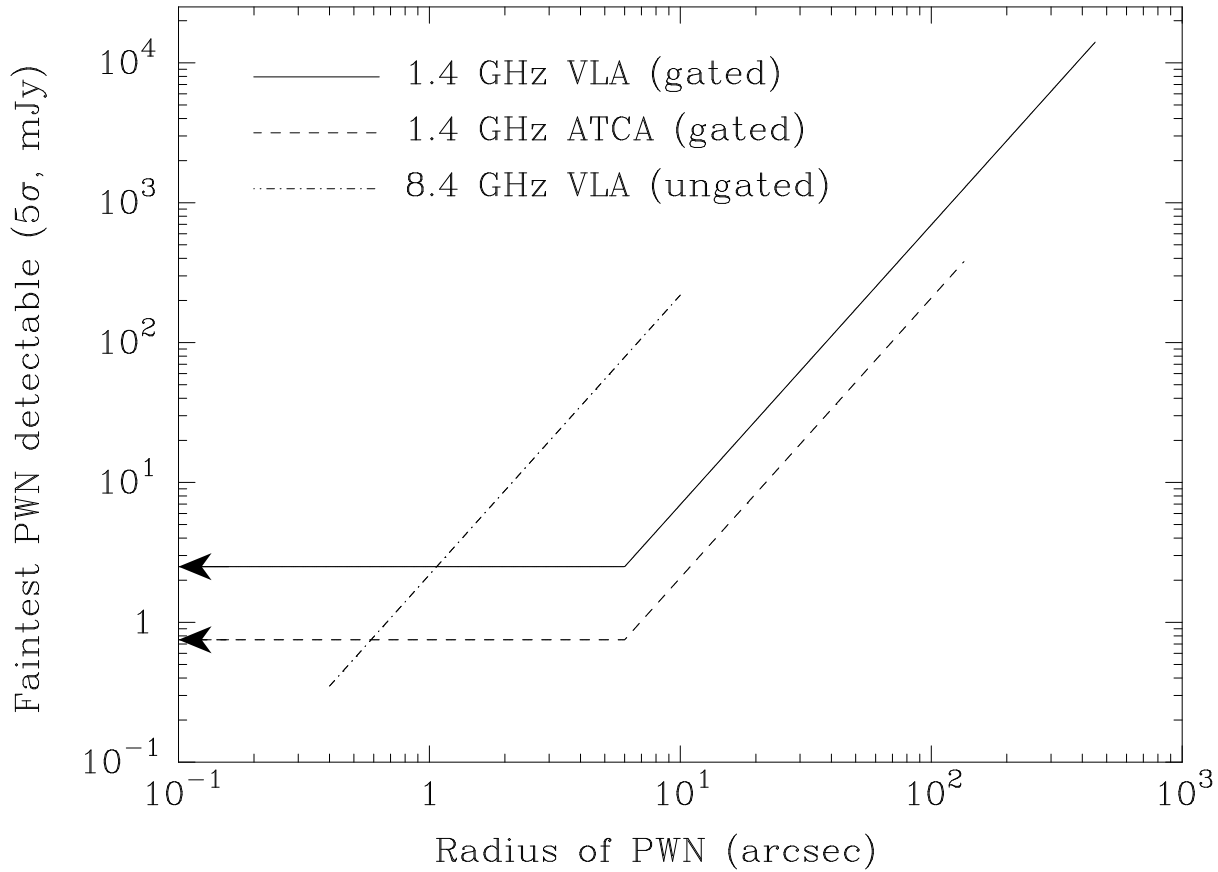}}
\caption{Comparison of 5$\sigma$ sensitivities of the 8.4~GHz PWN search of
FS97 with the 1.4~GHz observations reported here.
The observations of FS97 were not gated, and so generally are
not sensitive to PWN on scales smaller than the resolution limit.
For our VLA (ATCA)
observations, we have adopted a typical spatial resolution of $12''$
($12''$), a maximum spatial scale of $15'$ ($4\farcm5$), and a
1$\sigma$ sensitivity of 0.5 (0.15) mJy~beam$^{-1}$. A spectral
index $\alpha = -0.3$ ($S_\nu \propto \nu^{\alpha}$) has been assumed
in converting 8.4~GHz results to 1.4~GHz.}
\label{fig_sens}
\end{minipage} 
\end{figure*}

As discussed in \S\ref{sec_intro}, the parameters of the current survey
were chosen to improve on the sensitivity of previous surveys,
in particular that carried out
by FS97.  These factors are summarised in Fig~\ref{fig_sens}, where
the sensitivity of FS97's search is compared to that presented here.
Only for PWN with radii between $0\farcs4$ and $\sim1''$ is FS97's search
more sensitive than ours. Between $1''$ and $10''$, the current results
are considerably (up to 100 times) more sensitive, while at scales smaller
than $0\farcs4$ and larger than $10''$, our data probe a parameter space
to which FS97 were not sensitive at all.

The results reported in \S\ref{sec_results} showed that most of the
sources in our sample had no detectable PWN associated with them. The
exceptions were PSRs B0136+57, B1634--45 and B1706--16, for which
unresolved off-pulse sources were detected at the pulsar position. From
the current data, we are unable to conclusively determine whether this
emission corresponds to emission from the pulsar or from a compact
PWN. While we plan to investigate these sources further, for the
purposes of the present discussion we assume these observations to be
non-detections, but with a sensitivity corresponding to the flux density
of the off-pulse source (rather than to the noise in the
surrounding area of the image).

To quantify the significance of our non-detections,
we follow FS97 in characterising a radio PWN's integrated luminosity,
$L_R$, by
\begin{equation}
L_R = \epsilon \dot{E}~{\rm erg~s}^{-1},
\label{eqn_l_r}
\end{equation}
where  $\dot{E}$~erg~s$^{-1}$ is the associated pulsar's
spin-down luminosity, and $\epsilon$ is the fraction of $\dot{E}$
which goes into radio emission.

Assuming a typical PWN spectral index $\alpha = -0.3$ 
and integrating from 10~MHz to 100~GHz, the corresponding
1.4~GHz flux density is

\begin{equation}
S_{1.4} = 2.1\times10^{5}~\frac{\epsilon \dot{E}_{34}}{d^2}~{\rm mJy,}
\label{eqn_s20}
\end{equation}
where $d$~kpc is the distance to the pulsar and $\dot{E} = 10^{34}
\dot{E}_{34}$~erg~s$^{-1}$. We generally use distances from the pulsar
catalogue \cite{tmlc95}, derived either from a pulsar's
dispersion measure \cite{tc93} or from its kinematic distance based on
H~{\sc i} measurements (e.g. Frail \& Weisberg 1990\nocite{fw90}). 

If we do not detect a PWN in our observations, we can potentially put
an upper limit on $S_{1.4}$, and hence on $L_R$ and $\epsilon$. However,
as demonstrated in Fig~\ref{fig_sens}, these limits depend on the angular
size we expect for the PWN.  As previously discussed by FS97 and SGJ99,
PWN can in general be divided into two distinct classes (assuming that
the pulsar is not inside a SNR, which appears to be the case for all
the sources in our sample): those which are confined by the gas pressure
of the ambient ISM (``static PWN''), and those confined by ram-pressure
resulting from motion of the pulsar through the ISM (``bow-shock PWN'').

Let us first consider the case of a static PWN. The bubble in the ISM driven
by the pulsar wind will expand supersonically into the ambient medium,
producing a PWN of radius \cite{aro83}
\begin{equation}
R_{\rm static} = 0.14 \times \left( \frac{\dot{E}_{34} t^3_3}{n} 
\right)^{1/5}~{\rm pc},
\label{eqn_rstatic}
\end{equation}
where $n$~cm$^{-3}$ is the density of the ambient medium (assumed to be pure
hydrogen), and $t_3$~kyr is the period for which the pulsar has
been interacting with the ISM.
In further discussion we assume that this age is given by the pulsar's
characteristic age, $\tau_c \equiv P/2\dot{P}$.

However a PWN is only static while $\dot{R}_{\rm static} > V_{\rm PSR}$,
where $V_{\rm PSR}$~\kms\ is the pulsar's space velocity.
Re-arranging Equation~(3) of
FS97 (and correcting for a missing factor of $4\pi$ in their
results), we find that a PWN will be static for velocities and 
ambient densities for which:
\begin{equation}
n V_{\rm PSR}^5 \la 4 \times10^{9}~\dot{E}_{34}/t_3^2.
\label{eqn_nv5}
\end{equation}
When this condition is not met, the pulsar has
``overtaken'' its own static PWN, and a bow-shock PWN results. The resulting
PWN is much smaller than the static PWN, and has a size determined by a
balance between the pressure of the relativistic pulsar wind and the
ram-pressure resulting from the pulsar's motion,
\begin{equation}
R_{\mbox{\scriptsize bow-shock}} = 
0.63 \left(\frac{\dot{E}_{34}}{nV_{\rm PSR}^2}\right)^{1/2}~{\rm pc},
\label{eqn_rbow}
\end{equation}
where we have assumed that all the spin-down luminosity of the pulsar goes
into the wind, and that the radius of a bow-shock PWN is 1.5 times the
radius at which the ram and wind pressures balance \cite{vm88b}.

Thus for a given $n$ and $V_{\rm PSR}$, we can use Equation~(\ref{eqn_nv5})
to determine whether a PWN has a static or a bow-shock morphology, and from
this use either Equation~(\ref{eqn_rstatic}) or (\ref{eqn_rbow}) to
determine its radius, and hence its angular extent, $\theta_{\rm PWN}$.

For a pulsar-gated observation with a 1.4~GHz RMS sensitivity of
$\sigma$~mJy~beam$^{-1}$ at a resolution $\theta_\alpha '' \times
\theta_\delta''$, suppose no off-pulse source is detected at the 5$\sigma$
level.  There are three possible reasons for this non-detection: the PWN
is unresolved and below the point-source sensitivity of the observations,
the PWN is resolved and below the surface brightness sensitivity limit,
or the PWN is larger than the maximum spatial scale
to which we are sensitive. We now consider what
limits on $\epsilon$ can be derived  for each situation.

If $\theta_{\rm PWN}$ is smaller than the resolution limit, then $S_{1.4} <
5\sigma$ and from Equation~(\ref{eqn_s20}) the corresponding upper limit on
$\epsilon$ is
\begin{equation}
\epsilon < \epsilon_0 = 2.4\times10^{-5}~\sigma \left( \dot{E}_{34}/d^2 \right)^{-1}.
\label{eqn_eps_limit}
\end{equation}
Note that 
if an off-pulse point source of 1.4~GHz flux density $S_{\rm pt}$ is detected
at the position of the pulsar (as was the case for three of the pulsars
in our sample), then $5\sigma$ should be replaced by $S_{\rm pt}$
in this expression.

If the PWN is extended, but smaller than the largest spatial scale to which
our interferometer is sensitive, then
\begin{equation}
\epsilon < \frac{\epsilon_0 \theta_{\rm PWN}^2}{\theta_\alpha \theta_\delta},
\label{eqn_eps_limit2}
\end{equation}
where we have assumed a Gaussian profile for the PWN.
Finally, if the PWN is larger than our observations can detect,
we can put no limit on $\epsilon$.

FS97 have argued, on the basis of their non-detections, that the lack of
observable radio PWN around most pulsars implies $\epsilon \la 10^{-6}$, two
orders of magnitude less than for pulsars with detected radio PWN. However,
in making this calculation they assumed that the ambient density was
$n=1$~cm$^{-3}$. Through Equation~(\ref{eqn_nv5}), when combined with a
reasonable velocity, this implied that all the pulsars they observed were
powering bow-shock PWN, and the consequent high ram pressure ensured that
these sources would largely be unresolved by their $0\farcs8$ resolution.

However their assumed ambient density is not representative of our current
best understanding of the ISM. While the relative filling fractions are
uncertain, available evidence supports a multi-phase ISM of which 90\% by
volume is a combination of a warm medium of density $n = 0.3$~cm$^{-3}$ and
a hot ionised component of density $n = 0.003$~cm$^{-3}$ (see Ferri\`{e}re
1998a\nocite{fer98a} for a recent discussion and overview).  In the hot
low-density component, we still expect most PWN to be bow shocks, but the
ram pressure is greatly reduced and PWN will consequently be much more
extended. Thus many PWN in this low density medium will be larger than the
resolution limit of FS97, and through Equation~(\ref{eqn_eps_limit2}), the
limit on $\epsilon$ much less stringent than claimed.


In order to better consider the limits on $\epsilon$, we therefore carry out
the following calculation for 27 of the pulsars
in Table~\ref{tab_obs} (PSR~B1706--44 is excluded as it has
a candidate PWN), and additionally for the 4 pulsars discussed
by SGJ99. For each pulsar in our sample, we adopt possible
densities of $n=0.3$~cm$^{-3}$ and $n=0.003$~cm$^{-3}$.  Ten of the pulsars
in our sample have measured proper motions, and we use the corresponding 3D
space velocities determined by Cordes \& Chernoff \shortcite{cc98}; one
other pulsar (PSR~B1055--52) has had a scintillation velocity determined for
it \cite{jnk98}. For the remaining sources we set $V_{\rm PSR} = 380$~\kms,
corresponding to the mean pulsar velocity of the distribution of Cordes \&
Chernoff \shortcite{cc98}.  Using Equation~(\ref{eqn_nv5}) we determine
whether each corresponding PWN is bow-shock or static, and then consequently
determine $\theta_{\rm PWN}$. Upper limits on $\epsilon$ are then determined
from Equations~(\ref{eqn_eps_limit}) and (\ref{eqn_eps_limit2}),
and the results
given in Table~\ref{tab_limits}.
For the 16
sources in our sample which were also observed by FS97, we can make a
similar calculation based on their results (converting their data to 1.4~GHz
assuming a spectral index $\alpha = -0.3$); these revised values
of $\epsilon_{\rm max}$ are also listed in Table~\ref{tab_limits}.
Note that many PWN are unresolved for either choice of ambient density, 
and so have the same value of $\epsilon_{\rm max}$ in both cases.

\begin{table*}
\begin{minipage}{180mm}
\caption{Upper limits on $\epsilon = L_R/\dot{E}$ for PWN non-detections, 
including four non-detections from SGJ99.
Pulsars are sorted by decreasing $\dot{E}$; $\theta$ is the predicted
angular size of a PWN for a given density.
For upper limits on $\epsilon$ derived from the results of FS97,
U indicates that the predicted size of the PWN is too large to have
been detected by their observations, while ``$\ldots$'' indicates
that a particular pulsar was not part of their sample.}
\label{tab_limits}
\begin{tabular}{lcccccccccccc}
Pulsar & $\dot{E}_{34}$ & $t_3$ & $d$ & $V_{\rm PSR}$ &  &
\multicolumn{3}{c}{$n=0.3$~cm$^{-3}$} & & \multicolumn{3}{c}{$n=0.003$~cm$^{-3}$}  \\
       & (erg~s$^{-1}$)  & (kyr) & (kpc) & (\kms) & & $\theta$ ($''$) & 
\multicolumn{2}{c}{$\log_{10} \epsilon_{\rm max}$}
   & & $\theta$ ($''$) & \multicolumn{2}{c}{$\log_{10} \epsilon_{\rm max}$} \\
    &  & & & & & & (this paper) & (FS97) & & & (this paper) & (FS97) \\
B1823--13     &   280 &     21 &   4.1  & $\ldots$  &  &   5.01 &    --6.3  &
--5.4  &  &   50.1 &    --5.2  &     U      \\
J1105--6107   &   250 &     63 &   7.1  & $\ldots$  &  &   2.76 &    --5.8  & $\ldots$ & &    27.6 &    --5.6  &  $\ldots$  \\
B1046--58     &   200 &     20 &   3.0  & $\ldots$  &  &   5.86 &    --6.8  &  $\ldots$ &  &   58.6 &    --5.3  &  $\ldots$  \\
B1610--50     &   160 &      7 &   7.2  & $\ldots$  &  &   2.16 &    --5.8  &  $\ldots$ &  &   21.6 &    --5.3  &  $\ldots$  \\
B1727--33     &   120 &     26 &   4.2  & $\ldots$  &  &   3.19 &    --5.2  &    --5.3  &  &   31.9 &    --4.8  &     U      \\
B1930+22     &    75 &     40 &  12.1  & $\ldots$  &   &  0.88 &    --4.6  &    --5.4  &   &   8.8 &    --4.6  &    --3.4   \\
B0114+58     &    22 &    275 &   2.1  & $\ldots$  &   &  2.70 &    --5.6  &    --5.4  &   &  27.0 &    --5.1  &     U      \\
J1835--1106   &    18 &    127 &   3.1  & $\ldots$  &  &   1.69 &    --5.1  &  $\ldots$ &  &   16.9 &    --5.0  &  $\ldots$  \\
J0631+1036   &    17 &     44 &   6.6  & $\ldots$  &   &  0.78 &    --4.0  &  $\ldots$ &   &   7.8 &    --4.0  &  $\ldots$  \\
B0740--28     &    14 &    157 &   1.9  & 276       &  &   3.36 &    --5.5  &    --4.6  &  &   33.6 &    --4.9  &     U      \\
B1508--57     &    13 &    298 &  12.7  & $\ldots$  &  &   0.35 &    --4.2  &  $\ldots$ &  &    3.5 &    --4.2  &  $\ldots$  \\
B1356--60     &    12 &    319 &   5.9  & $\ldots$  &  &  0.72 &    --4.7  &  $\ldots$ &  &    7.2 &    --4.7  &  $\ldots$  \\
B1634--45     &     7.5 &    590 &    3.8  &  $\ldots$  & &    0.88 &    --5.1  &    --5.3  & &      8.8 &    --5.1  &    --3.3   \\
B0611+22     &     6.3 &     89 &    4.7  &  212       &  &   1.18 &    --4.5  &    --4.9  & &     11.8 &    --4.5  &    --2.9   \\
J0538+2817   &     4.9 &    619 &   1.8  &  $\ldots$  &   &  1.54 &    --5.2  &  $\ldots$ & &     15.4 &    --5.0  &  $\ldots$  \\
B1718--35     &     4.5 &    176 &   6.4  &  $\ldots$  &  &   0.41 &    --3.5  &  $\ldots$ & &      4.1 &    --3.5  &  $\ldots$  \\
B0355+54     &     4.5 &    563 &   2.1  &  210       &   &  2.29 &    --5.2  &    --4.8  & &     22.9 &    --4.7  &     U      \\
B0540+23     &     4.1 &    253 &   3.5  &  348       &   &  0.77 &    --4.5  &    --5.1  & &      7.7 &    --4.5  &    --3.3   \\
B1754--24     &     4.0 &    285 &   3.5 &  $\ldots$  &   &  0.71 &    --5.0  &  $\ldots$  & &      7.1 &    --5.0  &  $\ldots$  \\
B0656+14     &     3.8 &    111 &   0.8  &  331       &   &  3.63 &    --6.0  &    --5.2  &  &   36.3 &    --5.1  &     U      \\
B1719--37     &     3.3 &    345 &   2.5  & $\ldots$  &   &  0.89 &    --4.6  &    --5.4  &  &    8.9 &    --4.6  &    --3.4   \\
B1821--19     &     3.0 &    574 &   5.2  & $\ldots$  &   &  0.41 &    --4.4  &    --4.7  &  &    4.1 &    --4.4  &    --3.3   \\
B1055--52     &     3.0 &    535 &   1.5  & 440       &   &  1.21 &    --5.8  &  $\ldots$ &  &   12.1 &    --5.7  &  $\ldots$  \\
B2011+38     &     2.9 &    412 &  13.1  & $\ldots$  &    & 0.16 &    --2.9  &    --4.0  &   &   1.6 &    --2.9  &    --3.4   \\
B0136+57     &     2.1 &    403 &   2.9  & 340       &    & 0.69 &    --4.3  &    --4.6  &   &   6.9 &    --4.3  &    --3.3   \\
B1449--64     &     1.9 &   1035 &   1.8  & 337       &   &  1.04 &    --5.5  &  $\ldots$ &  &   10.4 &    --5.5  &  $\ldots$  \\
B1730--37     &     1.5 &    355 &   3.5  & $\ldots$  &   &  0.44 &    --3.6  &  $\ldots$ &  &    4.4 &    --3.6  &  $\ldots$  \\
B1933+16     &     0.51 &    947 &   7.9  &  996       &  &   0.04 &    --3.0  &  $\ldots$ & &     0.4 &    --3.0  &  $\ldots$  \\
B1706--16     &     0.12 &   1655 &   1.3  & 186       &  &   0.59 &    --3.5  &  $\ldots$ & &     5.9 &    --3.5  &  $\ldots$  \\
B0736--40     &     0.089 &   3805 &   2.1  & 377       & &    0.21 &    --3.8  &  $\ldots$ &&      2.1 &    --3.8  &  $\ldots$  \\
B2148+63     &     0.012 &  36640 &  13.6 &  $\ldots$  &  &   0.01 &    --0.7  &    --1.6  & &     0.1 &    --0.7  &    --1.6   \\
\end{tabular}
\end{minipage}
\end{table*}

We can compare our upper limits to values of $\epsilon$ for known PWN.
In Table~3 of FS97, $\epsilon$ is listed for the six pulsars then known to
have detected radio PWN, to which we add PWN which have since been
associated with PSR~J0537--6910 ($\epsilon = 5 \times 10^{-4}$; Lazendic \&
Dickel 1998\nocite{ld98}), PSR~B0906--49 ($\epsilon = 2\times 10^{-6}$; GSFJ98)
and PSR~J1811--1926 ($\epsilon < 2 \times 10^{-3}$; Morsi \& Reich
1987\nocite{mr87}; Torii \etal\ 1999\nocite{ttd+99}). 
For these nine sources, six measurements lie in the
reasonably narrow range $(1-5)\times10^{-4}$. Those sources lying outside
this range are PSR~B0833--45, for which it is unclear just what part of the
surrounding SNR is pulsar-powered, PSR~B0906--49, which GSFJ98 argue is
substantially different from other PWN, and PSR~J1811--1926, where the radio
PWN is faint and has poorly constrained properties.  Thus the data available
suggest that a ``typical'' detectable PWN has $\epsilon \approx 10^{-4}$.

FS97 argued that typical non-detections corresponded to $\epsilon_{\rm
max} \sim 2\times10^{-6}$, significantly less than for detected
PWN. However, it can be seen from the results in Table~\ref{tab_limits}
that for $n=0.003$~cm$^{-3}$ some PWN become too large to be detected
by their data, while for the remaining pulsars the limit rises to
$\epsilon_{\rm max} \sim 3\times10^{-4}$.  Thus we argue that the
observations of FS97 could have missed ``normal'' PWN around most
of their sample if most of these sources are in low density regions,
and hence their observations
do not constrain pulsars which lack detectable PWN to be any
different in their wind properties from those pulsars with observed PWN.

On the other hand, the current observations can potentially provide
constraining limits on $\epsilon$.  We first consider the six young and
energetic pulsars in our sample (i.e. the first six pulsars
in Table~\ref{tab_limits}). These pulsars are defined approximately
by $\dot{E}_{34} \ga 50$ and $t_3 \la 50$, properties similar to
those pulsars with detectable radio PWN. For either
assumed ISM density, our data constrain these pulsars to have upper limits
on $\epsilon$ in the range $(0.002-0.2)\times10^{-4}$, significantly less
than for pulsars with observed PWN. While this seems to imply genuinely
low values of $\epsilon$, for $n=0.003$~cm$^{-3}$, conditions are
such that these pulsars have only recently overtaken their static
nebulae. Lowering their velocities
slightly, arguing that their actual ages are less than their
characteristic ages, or accepting that realistically, the transition
from static to bow-shock PWN does not happen instantaneously, it seems
likely that these pulsars are still producing static nebulae,
whose extents are much larger than for bow-shocks. In this case, the
corresponding limits become $\epsilon_{\rm max} \gg 10^{-4}$, and are
not constraining.

We thus argue that our non-detections of PWN can be explained even
if all young and energetic pulsars have similar wind properties.
The difference between detectable and non-detectable PWN seems to be that
detected PWN are either in dense regions of the ISM or in SNRs, in which
there is sufficient external pressure to confine the pulsar wind and
produce an observable PWN. However, pulsars with no PWN are in the low
density phase of the ISM and so produce unobservable ``ghost remnants''
\cite{bopr73}.  With the exception of PSR~B0906--49 (GSFJ98\nocite{gsfj98}),
pulsars with observed PWN also have associated SNRs, while all those young
pulsars without PWN also have no associated SNRs. We indeed expect SNRs
to be undetectable in the hot component of the ISM \cite{ksbg80,gj95c},
consistent with our conclusion above that it is a low ambient density
which causes a PWN around a young pulsar to be undetectable.

A notable exception is PSR~B1757--24, which is associated with both a SNR
and a radio PWN despite having $n=0.003$~cm$^{-3}$ \cite{fk91,mkj+91}.
In this case, the pulsar is inferred to have a transverse space velocity
$\sim$1500~\kms\ \cite{fkw94}; this not only supplies the necessary
ram pressure to produce an observable PWN, but has caused the pulsar to
overtake the shell of the associated SNR, re-energising the remnant with
its passage. If it were not for the extreme velocity of the pulsar, neither the
PWN nor the SNR would be detectable, as expected in a low density region.

The relative numbers of young pulsars with detected and undetected SNRs/PWN
suggests an approximate filling fraction $\sim$50\% for the low density
component of the ISM. This is somewhat more than recent estimates of
15--20\% \cite{fer98b}, but can be explained by the fact that we expect
young pulsars to be preferentially located in low density regions produced
by the powerful winds of their progenitors and by previous supernovae
in the region.

\begin{figure*}
\begin{minipage}{160mm}
\centerline{\psfig{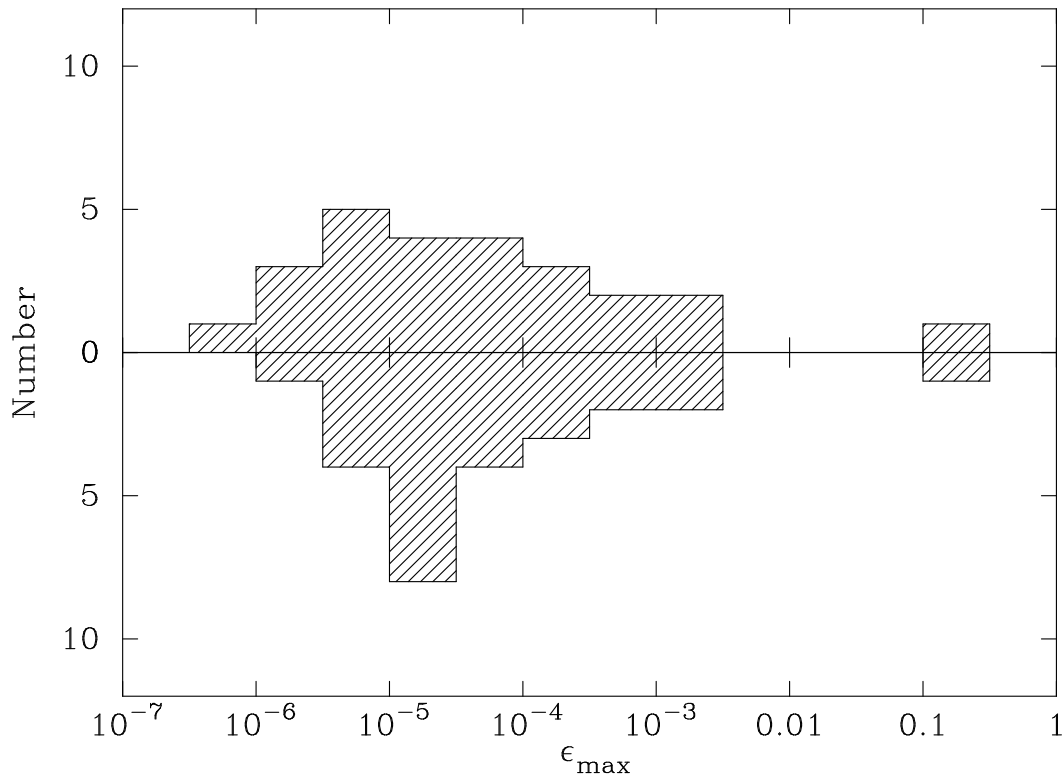}}
\caption{Upper limits on $\epsilon = L_R / \dot{E}$ from
pulsar-gated 1.4~GHz data in Table~\ref{tab_limits}; 
pulsars with $\dot{E}_{34} > 50$ have been excluded.
The upper panel corresponds to an assumed
ambient density $n=0.3$~cm$^{-3}$, while the lower
panel represents $n=0.003$~cm$^{-3}$.}
\label{fig_histo1}
\label{lastpage}
\end{minipage} 
\end{figure*}

The majority of pulsars in our sample are considerably less energetic
($\dot{E}_{34} < 50$) and older ($t_3 \ga 100$) than pulsars around
which radio PWN have been observed.  Values of $\epsilon_{\rm max}$ for
these pulsars are plotted in Fig~\ref{fig_histo1}, from which it can be
seen that for either choice of ambient density, the distribution peaks
around $\epsilon_{\rm max} \sim 10^{-5}$. These pulsars have all long
since overtaken their static PWN, and will have bow shock PWN for any
sensible choice of $n$ and $V_{\rm PSR}$. The resulting size of such a
PWN is not a strong function of $n$ or $V_{\rm PSR}$; to produce values
of $\epsilon_{\rm max} > 10^{-4}$, consistent with detected PWN, requires
uniformly low space velocities, $V_{\rm PSR} \la 150$~\kms, for these
pulsars. However, only $\sim$5\% of pulsars are thought to be traveling
at less than 150~\kms\ \cite{cc98}. While the ages and distances we
have used for these pulsars have their associated uncertainties, for
bow-shock PWN values of $\epsilon_{\rm max}$ are independent of age,
and distances would have to uniformly be increased by a factor of three
to shift the peak in $\epsilon_{\rm max}$ to a value in agreement with
that seen for detected PWN.  The lower value of $\epsilon$ we have
derived for older pulsars is thus a result quite robust to
the assumptions and uncertainties involved in its derivation, and we
therefore argue that these pulsars have winds which are genuinely at
least an order of magnitude less efficient at producing radio emission
than the winds of young and energetic pulsars.

FS97 consider various reasons why older pulsars might appear to have lower
values of $\epsilon$. Possibilities include:
\begin{enumerate}
\item that the PWN are resolved out by the observations;

\item that an increasing fraction of $\dot{E}$ goes into pulsed
X-rays and $\gamma$-rays (see Thompson \etal\ 1994\nocite{tab+94});

\item that their winds are dominated by Poynting flux rather than
relativistic particles;

\item that the injection spectrum of particles in the pulsar wind has
shifted to higher energies.
\end{enumerate}

Our observations can conclusively rule out alternative (i), as we can detect
PWN produced for almost all feasible values of $n$ and $V_{\rm PSR}$.
While we cannot distinguish between the remaining three possibilities, 
we note that of detected PWN, that associated with the oldest pulsar,
PSR~B0906--49, also has the lowest value of $\epsilon$ and the steepest
spectral index (GSFJ98). Since the spectral index of a PWN is directly related
to the spectrum of injected particles \cite{ps73}, this result tentatively
suggests that the efficiency of the wind in producing radio emission is
related to the injection spectrum, and that alternative (iv) might then best
explain the observations.

\section*{Conclusions} 

We have used pulsar-gating at 1.4~GHz to search for radio PWN around 27
pulsars. Our search was up to 100 times more sensitive than the only other
comparable survey, and was carried out on
spatial scales corresponding to a much wider range
of ambient densities and pulsar velocities. Including data from previous
work by SGJ99, non-detections towards 28 pulsars, plus inconclusive
results in three other cases, have allowed us to determine upper limits
on $\epsilon$, the fraction of a pulsar's spin-down luminosity which
goes into producing radio emission from a PWN.

We find that the data are consistent with virtually all young energetic
pulsars having $\epsilon \sim 10^{-4}$. The lack of PWN around $\sim$50\%
of young pulsars can be explained if they are in low ambient densities
(0.003~cm$^{-3}$), consistent with the absence of associated supernova
remnants around these sources.

For older pulsars, any reasonable choice of ambient density and pulsar
velocity results in upper limits on the wind efficiency $\epsilon <
10^{-5}$, ten times less than for young pulsars. Thus pulsars seem to
become less efficient at producing radio wind nebulae as they age;
we speculate that this result is due to the spectrum of injected
relativistic particles steepening in older pulsars.  This possibility can
be tested through X-ray observations towards such pulsars -- it is likely
that {\em Chandra}\ will make many new detections of X-ray PWN, and through
consequent imaging spectroscopy, we may finally be able to probe the
winds around older pulsars.

Of the $\sim$55 non-recycled pulsars with $\dot{E} > 3 
\times 10^{34}$~erg~s$^{-1}$, almost
all have now been searched for associated radio PWN down to a good
sensitivity. If those sources with no detectable PWN are indeed in low
density regions of the ISM, it seems unlikely that we will ever find radio
PWN around them with current telescopes.  For example, if PSR~B1046--58
is powering a static PWN with $\epsilon = 10^{-4}$, the resulting radio
nebula would be $20'$ across with a flux density at 1.4~GHz of 0.5~Jy.
To detect this source would require $\sigma = 0.3$~mJy~arcmin$^{-2}$,
which is generally below the confusion limit for instruments capable
of imaging sources this large.  We might have to wait for the  large
increase in sensitivity promised by the Square Kilometre Array in order
to make further progress.


\section*{Acknowledgements}

We are particularly grateful to Walter Brisken for his help with 
pulsar-gating at the VLA, Barry Clark and Miller Goss for their generous
re-scheduling of a failed observing run, and Andrew Lyne for supplying
timing solutions for many of the pulsars we observed.  We also  thank
Phillip Hicks for his assistance with the observations, Froney
Crawford for information on the pulse profile of PSR~B1634--45,
and Jon Arons,
Jim Cordes, Vicky Kaspi, Michael Pivovaroff and
Eric van der Swaluw for useful discussions and suggestions.
This research has made use of the data base of published pulse profiles
maintained by the European Pulsar Network, available at {\tt
http://www.mpifr-bonn.mpg.de/pulsar/data}, NASA's Astrophysics Data
System Abstract Service and of the SIMBAD database, operated at CDS,
Strasbourg, France.  The National Radio Astronomy Observatory is a
facility of the National Science Foundation operated under cooperative
agreement by Associated Universities, Inc. The Australia Telescope is
funded by the Commonwealth of Australia for operation as a National
Facility managed by CSIRO.  B.M.G. acknowledges the support of NASA
through Hubble Fellowship grant HF-01107.01-98A awarded by the Space
Telescope Science Institute, which is operated by the Association of
Universities for Research in Astronomy, Inc., for NASA under contract
NAS 5--26555.  B.W.S. is supported by NWO Spinoza grant 08-0 to E.P.J.
van den Heuvel.


\bibliographystyle{mn}
\bibliography{modrefs,psrrefs,crossrefs}

\begin{thebibliography}{{Johnston, Nicastro \& Koribalski }{1998}}

\bibitem[\protect\citename{Arons }{1983}]{aro83}
Arons~J., 1983, Nat, 302, 301

\bibitem[\protect\citename{Becker {\rm et~al. }}{1994}]{bwhz94}
Becker~R.~H., White~R.~L., Helfand~D.~J., Zoonematkermani~S., 1994, ApJS, 91,
  347

\bibitem[\protect\citename{Becker {\rm et~al. }}{1999}]{bkbm99}
Becker~W., Kawai~N., Brinkmann~W., Mignani~R., 1999, AA, 352, 532

\bibitem[\protect\citename{Blandford {\rm et~al. }}{1973}]{bopr73}
Blandford~R.~D., Ostriker~J.~P., Pacini~F., Rees~M.~J., 1973, JA\&A, 23, 145

\bibitem[\protect\citename{Brooks \& Whiteoak }{2000}]{bw99b}
Brooks~K.~J., Whiteoak~J.~B., 2000, MNRAS, submitted

\bibitem[\protect\citename{Cohen {\rm et~al. }}{1983}]{ccgm83}
Cohen~N.~L., Cotton~W.~D., Geldzahler~B.~J., Marcaide~J.~M., 1983, ApJ, 264,
  273

\bibitem[\protect\citename{Condon {\rm et~al. }}{1998}]{ccg+98}
Condon~J.~J., Cotton~W.~D., Greisen~E.~W., Yin~Q.~F., Perley~R.~A.,
  Taylor~G.~B., Broderick~J.~J., 1998, AJ, 115, 1693

\bibitem[\protect\citename{Cordes \& Chernoff }{1998}]{cc98}
Cordes~J.~M., Chernoff~D.~F., 1998, ApJ, 505, 315

\bibitem[\protect\citename{Cordova {\rm et~al. }}{1989}]{cmhm89}
Cordova~F.~A., Middleditch~J., Hjellming~R.~M., Mason~K.~O., 1989, ApJ, 345,
  451

\bibitem[\protect\citename{Ferri\`{e}re }{1998a}]{fer98a}
Ferri\`{e}re~K., 1998a, ApJ, 497, 759

\bibitem[\protect\citename{Ferri\`{e}re }{1998b}]{fer98b}
Ferri\`{e}re~K., 1998b, ApJ, 503, 700

\bibitem[\protect\citename{Frail \& Kulkarni }{1991}]{fk91}
Frail~D.~A., Kulkarni~S.~R., 1991, Nat, 352, 785

\bibitem[\protect\citename{Frail \& Scharringhausen }{1997}]{fs97}
Frail~D.~A., Scharringhausen~B.~R., 1997, ApJ, 480, 364 (FS97)

\bibitem[\protect\citename{Frail \& Weisberg }{1990}]{fw90}
Frail~D.~A., Weisberg~J.~M., 1990, AJ, 100, 743

\bibitem[\protect\citename{Frail, Goss \& Whiteoak }{1994}]{fgw94}
Frail~D.~A., Goss~W.~M., Whiteoak~J. B.~Z., 1994, ApJ, 437, 781

\bibitem[\protect\citename{Frail, Kassim \& Weiler }{1994}]{fkw94}
Frail~D.~A., Kassim~N.~E., Weiler~K.~W., 1994, AJ, 107, 1120

\bibitem[\protect\citename{Frater, Brooks \& Whiteoak }{1992}]{fbw92}
Frater~R.~H., Brooks~J.~W., Whiteoak~J.~B., 1992,
  J.\,Electr.\,Electron.\,Eng.\,Aust., 12, 103

\bibitem[\protect\citename{Gaensler \& Johnston }{1995}]{gj95c}
Gaensler~B.~M., Johnston~S., 1995, MNRAS, 277, 1243

\bibitem[\protect\citename{Gaensler {\rm et~al. }}{1998}]{gsfj98}
Gaensler~B.~M., Stappers~B.~W., Frail~D.~A., Johnston~S., 1998, ApJ, 499, L69
(GSFJ98)

\bibitem[\protect\citename{Gould \& Lyne }{1998}]{gl98}
Gould~D.~M., Lyne~A.~G., 1998, MNRAS, 301, 235

\bibitem[\protect\citename{Green {\rm et~al. }}{1999}]{gcl98}
Green~A.~J., Cram~L.~E., Large~M.~I., Ye~T., 1999, ApJS, 122, 207,
  (http://www.astrop.physics.usyd.edu.au/MGPS/)

\bibitem[\protect\citename{Greisen }{1996}]{gre96c}
Greisen~E., ed, 1996, The {\tt AIPS} Cookbook.
\newblock National Radio Astronomy Observatory, Charlottesville

\bibitem[\protect\citename{Johnston, Nicastro \& Koribalski }{1998}]{jnk98}
Johnston~S., Nicastro~L., Koribalski~B., 1998, MNRAS, 297, 108

\bibitem[\protect\citename{Johnston {\rm et~al. }}{1992}]{jlm+92}
Johnston~S., Lyne~A.~G., Manchester~R.~N., Kniffen~D.~A., D'Amico~N., Lim~J.,
  Ashworth~M., 1992, MNRAS, 255, 401

\bibitem[\protect\citename{Kafatos {\rm et~al. }}{1980}]{ksbg80}
Kafatos~M., Sofia~S., Bruhweiler~F., Gull~S., 1980, ApJ, 242, 294

\bibitem[\protect\citename{Kawai \& Tamura }{1996}]{kt96}
Kawai~N., Tamura~K., 1996, in Johnston~S., Walker~M.~A., Bailes~M., eds,
  Pulsars: Problems and Progress, {IAU} Colloquium 160.
\newblock Astronomical Society of the Pacific, San Francisco, p.~367

\bibitem[\protect\citename{Kennel \& Coroniti }{1984}]{kc84}
Kennel~C.~F., Coroniti~F.~V., 1984, ApJ, 283, 710

\bibitem[\protect\citename{Kijak {\rm et~al. }}{1998}]{kkwj98}
Kijak~J., Kramer~M., Wielebinski~R., Jessner~A., 1998, AAS, 127, 153

\bibitem[\protect\citename{Lazendic \& Dickel }{1998}]{ld98}
Lazendic~J.~S., Dickel~J.~R., 1998, Mem. S. A. It., 69, 843

\bibitem[\protect\citename{Lockman, Pisano \& Howard }{1996}]{lph96}
Lockman~F.~J., Pisano~D.~J., Howard~G.~J., 1996, ApJ, 472, 173

\bibitem[\protect\citename{Manchester \& Johnston }{1995}]{mj95}
Manchester~R.~N., Johnston~S., 1995, ApJ, 441, L65

\bibitem[\protect\citename{Manchester {\rm et~al. }}{1991}]{mkj+91}
Manchester~R.~N., Kaspi~V.~M., Johnston~S., Lyne~A.~G., D'Amico~N., 1991,
  MNRAS, 253, 7P

\bibitem[\protect\citename{Michel }{1982}]{mic82}
Michel~F.~C., 1982, Reviews of Modern Physics, 54, 1

\bibitem[\protect\citename{Morsi \& Reich }{1987}]{mr87}
Morsi~H.~W., Reich~W., 1987, A\&AS, 71, 189

\bibitem[\protect\citename{Napier, Thompson \& Ekers }{1983}]{nte83}
Napier~P.~J., Thompson~A.~R., Ekers~R.~D., 1983, Proc.\,I.\,E.\,E.\,E., 71,
  1295

\bibitem[\protect\citename{Navarro {\rm et~al. }}{1995}]{nbf+95}
Navarro~J., de~Bruyn~G., Frail~D., Kulkarni~S.~R., Lyne~A.~G., 1995, ApJ, 455,
  L55

\bibitem[\protect\citename{Pacini \& Salvati }{1973}]{ps73}
Pacini~F., Salvati~M., 1973, ApJ, 186, 249

\bibitem[\protect\citename{Rees \& Gunn }{1974}]{rg74}
Rees~M.~J., Gunn~J.~E., 1974, MNRAS, 167, 1

\bibitem[\protect\citename{Sault \& Killeen }{1998}]{sk98}
Sault~R.~J., Killeen~N. E.~B., 1998, The Miriad User's Guide.
\newblock Australia Telescope National Facility, Sydney,
  (http://www.atnf.csiro.au/computing/software/miriad/)

\bibitem[\protect\citename{Sch\"{o}nhardt }{1974}]{sch74}
Sch\"{o}nhardt~R.~E., 1974, AA, 35, 13

\bibitem[\protect\citename{Stappers, Gaensler \& Johnston }{1999}]{sgj99}
Stappers~B.~W., Gaensler~B.~M., Johnston~S., 1999, MNRAS, 308, 609 (SGJ99)

\bibitem[\protect\citename{Taylor \& Cordes }{1993}]{tc93}
Taylor~J.~H., Cordes~J.~M., 1993, ApJ, 411, 674

\bibitem[\protect\citename{Taylor, Manchester \& Lyne }{1993}]{tml93}
Taylor~J.~H., Manchester~R.~N., Lyne~A.~G., 1993, ApJS, 88, 529

\bibitem[\protect\citename{Taylor {\rm et~al. }}{1995}]{tmlc95}
Taylor~J.~H., Manchester~R.~N., Lyne~A.~G., Camilo~F.
\newblock 1995.
\newblock Unpublished (available at ftp://pulsar.princeton.edu/pub/catalog)

\bibitem[\protect\citename{Thompson {\rm et~al. }}{1994}]{tab+94}
Thompson~D.~J. {\rm et~al.}, 1994, ApJ, 436, 229

\bibitem[\protect\citename{Torii {\rm et~al. }}{1999}]{ttd+99}
Torii~K., Tsunemi~H., Dotani~T., Mitsuda~K., Kawai~N., Kinugasa~K., Saito~Y.,
  Shibata~S., 1999, ApJ, 523, 69

\bibitem[\protect\citename{van Buren \& McCray }{1988}]{vm88b}
van Buren~D., McCray~R., 1988, ApJ, 329, L93

\bibitem[\protect\citename{Wang {\rm et~al. }}{2000}]{wmp+99}
Wang~N., Manchester~R.~N., Pace~R., Bailes~M., Kaspi~V.~M., Stappers~B.~W.,
  Lyne~A.~G., 2000, MNRAS, submitted.

\bibitem[\protect\citename{Weiler \& Panagia }{1978}]{wp78}
Weiler~K.~W., Panagia~N., 1978, AA, 70, 419

\bibitem[\protect\citename{Weiler, Goss \& Schwarz }{1974}]{wgs74}
Weiler~K.~W., Goss~W.~M., Schwarz~U.~J., 1974, AA, 35, 473

\end{thebibliography}

\clearpage




\end{document}